\documentclass[11pt,a4paper]{article}
\usepackage{jcappub}
\usepackage{amsfonts}
\usepackage{subfigure}

\renewcommand{\thesubfigure}{\alph{subfigure}}

\newcommand{\be}{\begin{equation}}
\newcommand{\ee}{\end{equation}}
\newcommand{\bea}{\begin{eqnarray}}
\newcommand{\eea}{\end{eqnarray}}

\begin{document}



\title{Turbulence patterns and neutrino flavor transitions in high-resolution supernova models}

\author[a]{Enrico Borriello,}
\author[b]{Sovan Chakraborty,}
\author[c]{Hans-Thomas Janka,}
\author[d]{Eligio~Lisi,}
\author[a]{Alessandro Mirizzi}

\affiliation[a]{II.~Institut f\"ur Theoretische Physik, Universit\"at
Hamburg\\ Luruper Chaussee 149, D-22761 Hamburg, Germany}
\affiliation[b]{Max-Planck-Institut f\"ur Physik
(Werner-Heisenberg-Institut)\\
 F\"ohringer Ring 6, D-80805 M\"unchen, Germany}
\affiliation[c]{Max-Planck-Institut f\"ur Astrophysik, Karl-Schwarzschild-Str. 1, 85748 Garching, Germany}
\affiliation[d]{Istituto Nazionale di Fisica Nucleare, Sezione di Bari, 
               Via Orabona 4, 70126 Bari, Italy}

\emailAdd{enrico.borriello@desy.de, sovan@mppmu.mpg.de, thj@mpa-garching.mpg.de}  
\emailAdd{eligio.lisi@ba.infn.it, alessandro.mirizzi@desy.de}

\abstract{
During the shock-wave propagation in a core-collapse supernova (SN), 
matter turbulence may affect neutrino flavor conversion probabilities. 
Such effects have been usually studied by adding parametrized small-scale random fluctuations (with arbitrary amplitude) 
on top of coarse, spherically symmetric matter density profiles. Recently, however, two-dimensional (2D) SN models have 
reached a space resolution high enough to directly trace anisotropic density profiles, down to scales smaller than the 
typical neutrino oscillation length. In this context, 
we analyze the statistical properties of a large set of SN matter density profiles obtained in a high-resolution 2D simulation, 
focusing on a post-bounce time (2~s) suited to study shock-wave effects on neutrino propagation 
on scales as small as O(100) km and possibly below. We clearly find the imprint
of a broken (Kolmogorov-Kraichnan) power-law structure, as generically expected in 2D turbulence spectra. We then compute     
the flavor evolution of SN neutrinos along representative realizations of the turbulent matter density profiles, and 
observe no or modest damping of the neutrino crossing probabilities  on their way through the shock wave. 
In order to check the effect of possibly unresolved fluctuations at scales below O(100) km, we also apply 
a randomization procedure anchored to the power spectrum calculated from the  simulation, and find consistent results within $\pm 1\sigma$ fluctuations. These results show the importance of anchoring turbulence effects on SN neutrinos 
to realistic, fine-grained SN models.

}
\maketitle

\section{Introduction}
The neutrino flux emitted during a core-collapse supernova
(SN) explosion represents an intriguing astrophysical case where flavor
transformations exhibit a strong sensitivity  both 
on the core-collapse dynamics and on unknown neutrino properties (such as the mass hierarchy). The relevant
phenomena depend strongly on the background fermion density and thus on the radial distance $r$ from
the SN center.

In the deepest
SN regions [$r \lesssim {\mathcal O} (10^2-10^3)$~km] the neutrino density is so high that 
neutrino-neutrino interactions become dominant~\cite{Pantaleone:1992eq,Qian:1994wh} and can induce 
a neutrino refractive index responsible for collective flavor conversions; see, e.g.,
\cite{Duan:2005cp,Duan:2006an,Hannestad:2006nj,Fogli:2007bk} and the recent review  \cite{Duan:2010bg}.
The main observable features of such flavor transitions consist in
an exchange of the spectrum of the electron species $\nu_e$ (${\bar\nu}_e$) with
the non-electron ones  $\nu_x$ (${\bar\nu}_x$) in certain energy intervals, giving
rise to 
so-called spectral
 ``swaps'' and ``splits''~\cite{Duan:2007bt,Dasgupta:2009mg,Mirizzi:2013wda}.

 While neutrinos propagate towards larger radii, their collective conversions get suppressed as their distribution becomes increasingly radially beamed so that eventually the flavor conversions are dominated by the ordinary matter.
In this context, 
many studies have analyzed 
possible signatures of the
Mikheyev-Smirnov-Wolfenstein (MSW) matter effect~\cite{Matt,Dighe:1999bi}
in increasingly realistic and detailed SN density profiles. For example,
since the neutrino conversion probabilities depend  on the
time-dependent matter profile, a high-statistics SN neutrino
observation may also reveal signatures
of shock-wave propagation at late times (i.e. $t \gtrsim 2$~s after the core-bounce)~\cite{Schirato:2002tg,Fogli:2003dw,Fogli:2004ff,Tomas:2004gr,Gava:2009pj}.
The transient violation of the adiabaticity condition when neutrinos cross the shock-front 
would then emerge as 
an observable modulation of the neutrino signal.
This signature would be particularly interesting to follow in ``real-time'' the shock-wave propagation, as well as
to probe the neutrino mass hierarchy. 
 
A realistic characterization of matter effects during neutrino propagation across the SN shock wave 
must also take into account stochastic density fluctuations and inhomogeneities
of various magnitudes and
correlation lengths in the ejecta layer in the wake of the shock front. 
These anisotropies and perturbations are a result of hydrodynamic 
instabilities between the proto-neutron star and the SN shock 
during the very early stages of the SN explosion. They lead to 
large-scale explosion asymmetries and turbulence in a dense shell
of shock-accelerated ejecta, which subsequently also seed secondary 
instabilities in the outer shells of the exploding star (e.g., 
Refs.~\cite{Kifonidis:2003,Kifonidis:2006,Scheck:2006,Hammer:2010,Arcones:2011,Mueller:2012,Wongwathanarat:2013}).

Neutrino flavor conversions in a stochastic matter background have been subject of intense
investigations both in a general 
context~\cite{Schaefer:1987fr,Sawyer:1990tw,Loreti:1994ry,Nunokawa:1996qu,Balantekin:1996pp,Burgess:1996mz,torrente}
 and specifically in relation to SN neutrinos~\cite{Loreti:1995ae,Fogli:2006xy,Choubey:2007ga,Friedland:2006ta,Kneller:2010sc,random,Kneller:2013ska,Lund:2013uta}. 
In general, one expects that
stochastic matter fluctuations of sufficiently large amplitude may suppress flavor
conversions and eventually lead to flavor equilibration between 
electron and non-electron species. Therefore, the spectral properties of the fluctuations
are very important for understanding the neutrino signal emerging from a core-collapse SN.

Ideally, the study of matter density effects on neutrinos propagating through a SN shock wave
should be based on input matter profiles derived from advanced and preferably non-isotropic SN models,
with space resolution smaller than typical oscillation lengths. Such models should be evolved to the late times (a few seconds)
relevant for shock-wave effects on neutrinos. In the absence of such ideal and rather demanding requirements,
many phenomenological studies have characterized the turbulence in terms of ``plausible'' disturbances, 
with arbitrary or tunable amplitudes, superimposed to shock-wave profiles derived by coarse, spherically symmetric SN models.

In this context, several relevant works aimed at characterizing and modeling the fluctuation spectra beyond earlier approximations~\cite{Fogli:2006xy,Choubey:2007ga} of delta-correlated noise.
\footnote{Neutrino propagation through non delta-correlated turbulence was also studied in~\cite{Benatti:2004hn}.} 
In particular, in~\cite{Friedland:2006ta} a general description of the turbulence was proposed, by modeling
 the density fluctuations via a Kolmogorov-type power spectrum.  
This model would allow to include matter fluctuations on all
the scales between the shock-front radius and the cut-off scale
by viscous dissipation.
A ``shadow effect'' on the shock-wave signature was estimated  for sufficiently large fluctuations.
In~\cite{Kneller:2010sc}  it  was described how to generate
a fluctuation power-spectrum on top of an undisturbed density profile through the 
``randomization method''~\cite{random}. 
Parametric studies were recently   performed to investigate the dependence of the damping effects on the amplitude
of the fluctuations~\cite{Kneller:2013ska}, and the impact on signatures related to the shock wave and to collective oscillations~\cite{Lund:2013uta}.

We build upon previous works on this topic, by anchoring the turbulence spectra 
to high-resolution SN simulations which have 
become available in recent years, although still in two-dimensional (2D) approximation. 
In particular, 2D simulations extending from core bounce to the late
post-bounce times relevant for neutrino propagation 
have become available by works of the Garching group~\cite{Kifonidis:2003,Kifonidis:2006}. 
Very interestingly, the power spectra of radial density profiles from such simulations
show the typical imprint expected from 2D turbulence.%
\footnote{A preliminary characterization  of turbulence spectra for these 2D simulations was presented in~\cite{rashba}.} 
The aim of the current study is thus to perform a self-consistent study of 
the SN matter turbulence by using such 2D simulation inputs, eventually improved with randomized fluctuations 
at small scales below O(100) km, which are not necessarily captured by the simulations themselves at large radii.

We also mention that, although first three-dimensional (3D)
simulations covering the relevant SN stages some seconds after core bounce have been carried out more recently~\cite{Hammer:2010,Mueller:2012,Wongwathanarat:2013}, the resolution of these models is still much coarser than those of the 2D calculations used for the analysis in our paper. 
Moreover, due to the lack of sufficient resolution and thus the
overestimation of numerical dissipation, the expected Kolmogorov
power law~\cite{Dolence:2013,Couch:2013b} cannot be recovered by preliminary analyses of
the corresponding fluctuation spectra.
For these reasons we shall limit ourselves to the 2D simulations of \cite{Kifonidis:2003,Kifonidis:2006} in this work.

Our results are presented as follows. 
In Sec.~2 we describe the  input matter profiles as taken from our reference 2D SN model.
In Sec.~3 we perform a statistical study of the matter power spectra of the density fluctuations in the different
radial directions. We find that these  spectra can be described by   broken power-laws, with a  ``typical'' structure which is actually
the imprint of two-dimensional turbulence.  
In Sec.~4 we use some representative turbulent matter profiles 
to calculate the neutrino crossing probabilities. We find that, by taking our reference SN profiles at face value,
turbulence provides no or modest damping of flavor transitions. 
As a relevant consequence, the shock-wave signature  on the neutrino crossing probabilities should hopefully remain
observable in the neutrino signal from a future galactic SN explosion. 
In Sec.~5  we investigate the effect of fluctuations on scales smaller than the typical resolution of the simulations, by generating them with a randomization method anchored to larger scales. We find that such small scale fluctuations may provide limited additional damping in some cases, but do not qualitatively alter the main picture emerging from the 2D simulations.
Finally, in Sec.~6 we discuss and summarize our findings.


\section{Our numerical SN model in 2D}

Our investigations are based on results of the 2D SN explosion simulations
of a nonrotating, 15\,$M_\odot$ star (a blue supergiant progenitor model
of Supernova~1987A from \cite{Woosley:1988at}) 
described in detail in Ref.~\cite{Kifonidis:2006}. 
The explosions via the neutrino-heating mechanism were launched by  
suitably chosen neutrino luminosities from the high-density core of
the nascent neutron star, and the evolution was continuously modelled
in 2D from a few milliseconds after core bounce until 20,000 seconds
later. The adaptive mesh refinement code AMRA was used, which allowed
to temporarily achieve a radial resolution equivalent to covering the
entire star (out to its surface radius of roughly $3\times 10^{12}$\,cm)
with 2.6 million equidistant radial zones, although only 3072 radial
zones were effectively used on the finest resolution level of the
adaptive numerical grid. 
A linear interpolation was used among the different zones. During the simulation all hydrodynamical quantities were interpolated with a higher order (PPM = piecewise parabolic) scheme within the grid cells.
The effective number of 
angular zones on the finest grid level was 768. 
The exact use of the 2D grid was described in~\cite{Kifonidis:2003}
to which we  address the reader to that reference for further details.
In the following we
made use of the model run b18b of~\cite{Kifonidis:2006}, whose 
explosion energy had a value of $10^{51}$\,erg.

For the further analysis, the computational output in the relevant
(dynamically changing) radial region was mapped to a uniform spatial 
polar grid of 3072 radial and 768 angular zones. From these data
of the two-dimensional SN model at different evolution times we
extracted the matter density profiles  on $N=768$ radial directions,
having an angular separation of $\Delta \theta = 180^\circ/768 \sim 0.234^\circ$.
On each direction the radial resolution is as fine as 
$\delta r \sim {\mathcal O}(10)$~km at late times ($t \sim \textrm{few}$~s). 

\begin{figure}[!t]
\centering
\includegraphics[width=0.45\columnwidth]{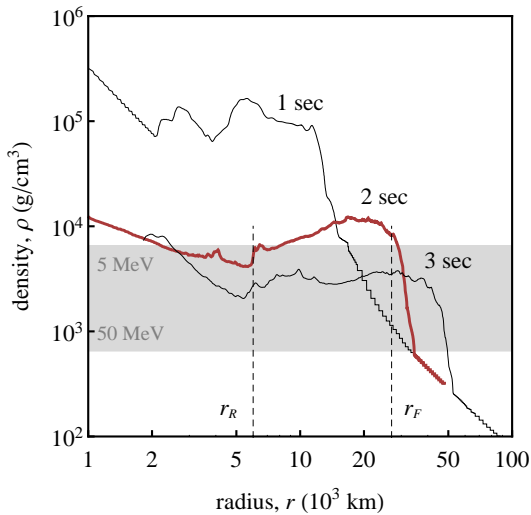}
\caption{Radial evolution of the angle-averaged 
SN matter density profile in our numerical calculation at the indicated post-bounce times.
The positions of the forward-shock shock ($r_F$) and of the 
reverse-shock ($r_R$) at $t=2$~s are roughly indicated. It is important
to note that because of the global nonsphericity of the explosion
(cf.~Fig.~\ref{fig2}) both forward and reverse shocks appear
smeared out in the angle-averaged profiles more widely than caused by
the finite numerical resolution. The horizontal band corresponds to
the density range of resonant neutrino oscillations in matter with 
the atmospheric mass
difference. The width of the band reflects the energy
range  $E\in[5,50]$~MeV.} \label{fig1}
\end{figure}

\begin{figure}[!t]
\centering
\includegraphics[width=0.7\columnwidth]{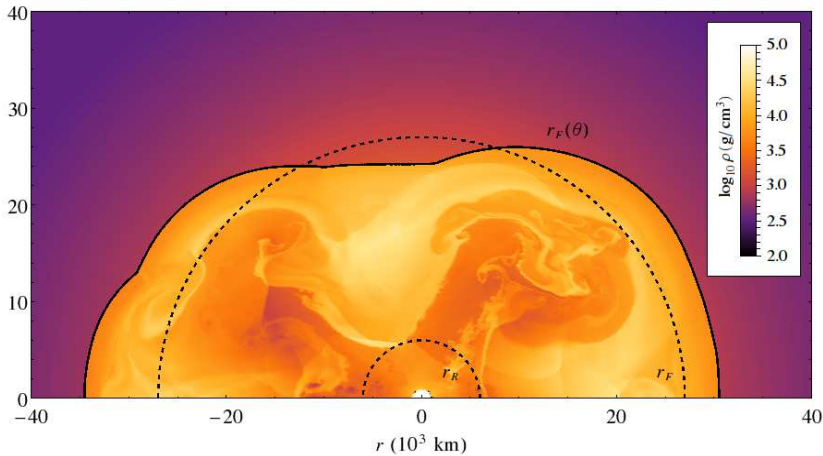}
\caption{Two-dimensional density plot of the SN simulation at $t=2$~sec
after supernova shock formation. The thick solid line marks the angle-dependent
shock front position. The outer (inner) dashed line marks the average forward shock
(reverse shock) front radius.} \label{fig2}
\end{figure}

In Fig.~\ref{fig1} we show the radial evolution of  angle-averaged density profiles,
\begin{equation}
\rho(r) =\frac{1}{N}\sum_{i=1}^N \rho_i(r) \,\ ,
\end{equation}
at different post-bounce times (1, 2 and 3~seconds).
One recognizes that  the SN shock wave at $r_F$, 
while propagating outwards at supersonic
speed, is followed by a dense shell delimited by two flow
discontinuities, the forward shock front (at $r_F$) on the one side and
an inner one, called \emph{reverse} shock front at position $r_R$ with
$r_R < r_F$. The forward shock expands into the hydrostatic density
profile of the progenitor star at $r > r_F$. The reverse shock is a
consequence of the fast neutrino-driven wind (a low-density, high-entropy
outflow blown off the nascent neutron star's surface by continuous
neutrino heating) colliding with the slower, shock-accelerated SN ejecta
(a detailed discussion can be found in~\cite{Arcones:2007,Arcones:2011}).
The layer between both shocks contains the anisotropic and strongly 
perturbed neutrino-heated ejecta, which
were subject to hydrodynamic instabilities before
the onset of the explosion. The initial asymmetries continue to grow and
trigger secondary mixing and fragmentation processes in this shell,
leading to large inhomogeneities of the density distribution on all scales
from the global one down to the resolution limit of the numerical 
simulation (see Fig.~\ref{fig2}).

The MSW effect causes a resonant level crossing in neutrino oscillations  when~\cite{Kuo:1989qe}
\begin{equation}
\frac{\Delta m^2_{\rm atm}}{2E} \simeq \pm \sqrt{2} G_F Y_e \rho \,\ ,
\label{eq:res}
\end{equation}
where $\Delta m^2_{\rm atm} =m_3^2 - m_{1,2}^2$ is the atmospheric mass-square difference,
while the  
  plus and minus sign at the right-hand-side refers to neutrinos $\nu$ and antineutrinos
$\bar\nu$, respectively.~\footnote{We explicitly checked that   the dependence of the SN resonant flavor conversions on
the subleading ``solar'' squared mass gap $\Delta m^2_{\rm sol}$ and mixing angle $\theta_{12}$ is negligible 
for our SN model.}
 Therefore, for normal mass hierarchy 
(NH, $\Delta m^2_{\rm atm} >0$) the resonance occurs in the $\nu$ channel, while for inverted
mass hierarchy (IH, $\Delta m^2_{\rm atm} <0$) it happens for $\bar\nu$. 
The electron fraction $Y_e$ in Eq.~(\ref{eq:res}) is $\sim 0.5$ in the region relevant
for the MSW effect.

With the horizontal band in Fig.~1 we indicate the density region corresponding to resonant
neutrino oscillations for a typical SN neutrino energy range
$E \in [5,50]$~MeV and for $\Delta m^2_{\rm atm} = 2.0\times 10^{-3}$~eV$^2$ (used hereafter), close to the current best-fit value~\cite{Fogli:2012ua}.
It turns out that the case of $t=2$~sec in Fig.~1 represents  
the best time snapshot  to have resonant flavor conversions for all the relevant neutrino energies, 
both across the sharp forward shock front and along its turbulent wake (back to the reverse shock front). 
The $t=2$~sec profile also happens to have the best radial resolution ($\delta r=15.3$~km)
as compared to the others at lower or higher post-bounce time (where $\delta r=46.3$~km or higher). 
Therefore, in the following we will take this case as benchmark for our study.

The reason for showing first the angle-averaged radial profile is that, in our reference 2D simulations, 
the matter density in different directions exhibits large variations with respect to the average, 
which in some cases make it difficult to recognize the forward+reverse-shock structure.
In this context we present in Fig.~2 the two-dimensional density plot for our reference SN simulation at $t=2$~s. 
The forward shock position $r_F(\theta)$ in different directions
(connected by the solid tick curve) can be rather different from the
average one at  $r_F = 2.7 \times 10^4$~km (dashed curve). Moreover, not at all angles the 
reverse shock is clearly visible 
in Fig.~2, although it emerges at $r_R \sim 6\times 10^3$~km
(inner dashed curve in Fig.~2) after angular averaging (see Fig.~1). Like the forward
shock, the reverse shock is highly aspherical. It is strong at the head of slowly expanding,
high-density downflows, whereas it is weak in regions where high-entropy, low-density
bubbles rise faster by acceleration through buoyancy forces. The average position is determined
by the massive, dense downdrafts.
For later purpose, we shall take an average range of $\Delta r=r_F-r_R=2.1\times 10^4$~km
as representative of the radial interval between the two shock waves, where turbulence is important.

In order to show the large density variations and differences radial profiles can exhibit,
we plot the radial matter-density profiles along three representative 
directions in Fig.~3.
While the matter profiles in the direction $i=9$ (corresponding to $\theta= -88.0^\circ$) in the left panel and at
 $i=466$ ($\theta= 19.2^\circ$) in the right panel are rather smooth, the profile at $i=177$ ($\theta= -48.5^\circ$) in the middle panel 
displays a ragged behavior downstream of the forward shock (at $r < r_F$),
with many contact discontinuities due to intense turbulence. 
The horizontal bands represent the MSW resonant regions for the same energy range as in Fig.~1.%
\footnote{
We remark that the shock fronts from the SN simulations have been steepened ``by hand,''
in order to correct artifacts due to finite resolution. This correction is needed to properly account for
strong violations of adiabaticity in  neutrino flavor 
conversions across the shock discontinuity.
} 
 We realize that, in general, multiple resonances  can arise along the matter-density profile
behind the SN shock front (i.e., at $r < r_F$). 
Since the resonance pattern is quite different in these three cases, one might expect
a different impact of turbulence on  the conversion probabilities. However, as we shall see in Sec.~4, the impact of fluctuations
is actually quite modest, although it may somewhat increase by adding further small-scale fluctuations via
randomization (see Sec.~5).

\begin{figure}[!t]
\centering
\includegraphics[width=0.325\columnwidth]{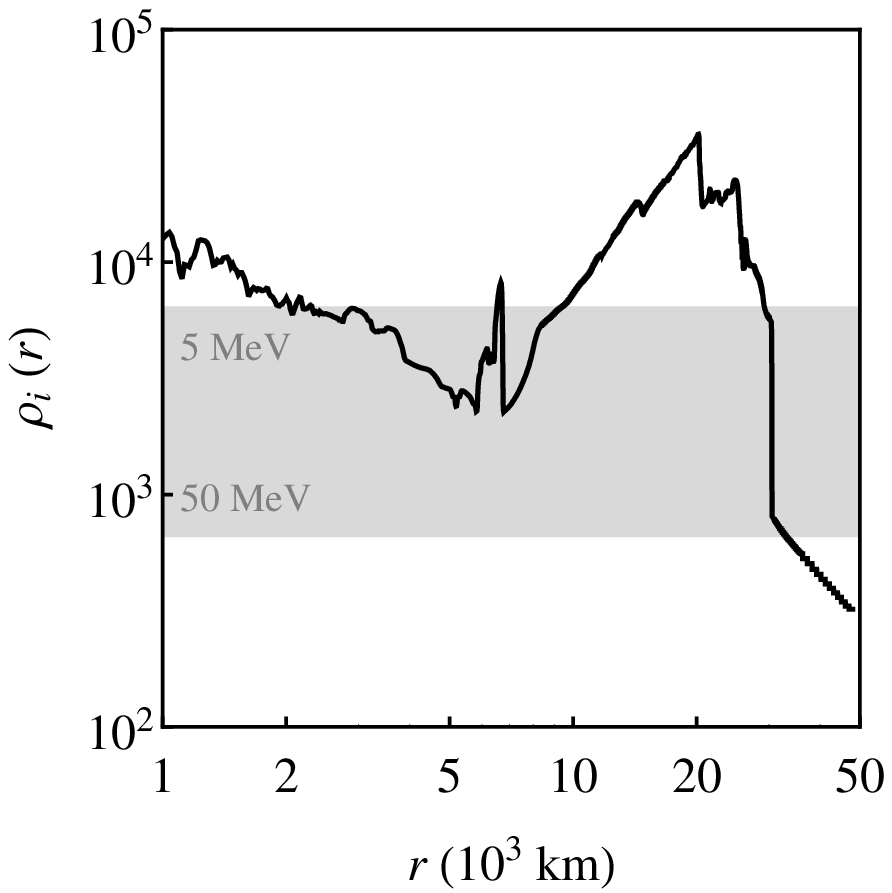}
\includegraphics[width=0.325\columnwidth]{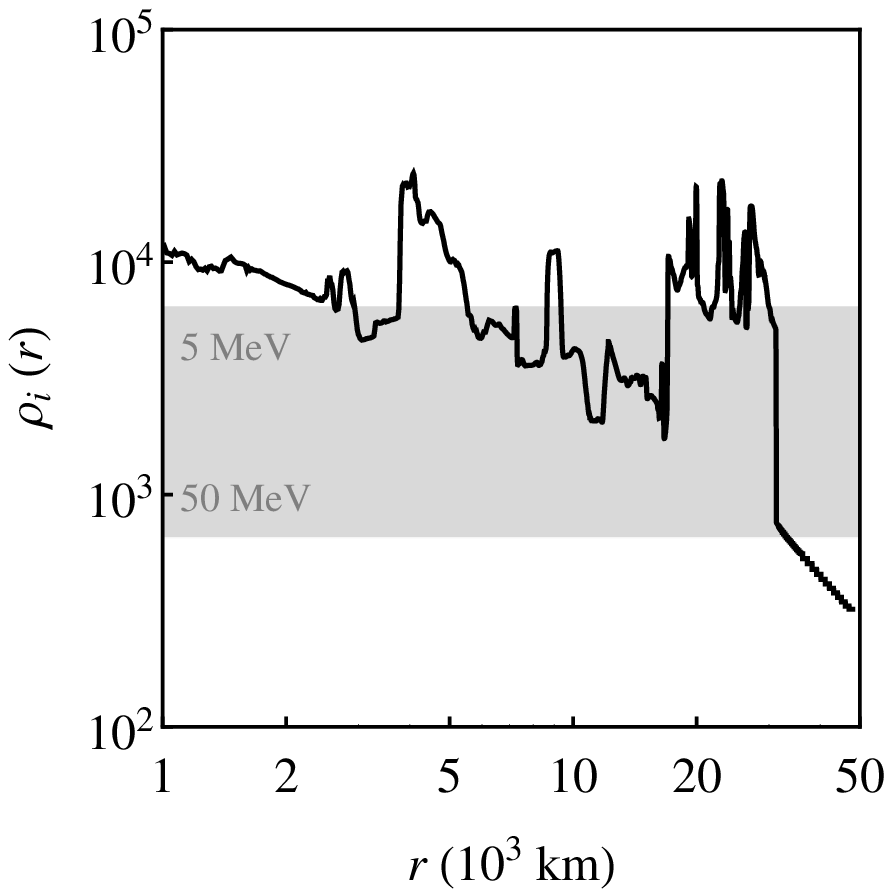}
\includegraphics[width=0.325\columnwidth]{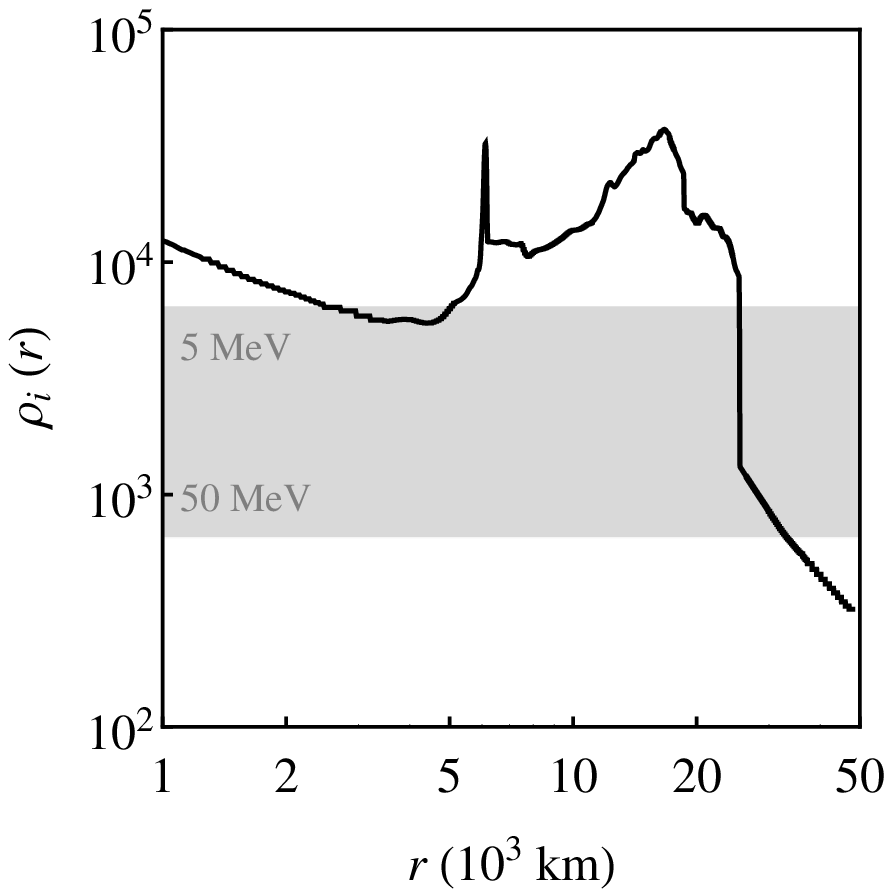}
\caption{SN matter density profiles for three representative directions
at post-bounce time $t=2$~s. The horizontal band represents the resonant region for
the same energy range of Fig.~1.  
} \label{fig3}
\end{figure}

\section{Power-spectrum of the matter density fluctuations} 

While there is little to be learned from a mere collection of density profiles $\rho_i$ as in Fig. \ref{fig3}
--as a big variety of different cases can be observed-- it turns out that the power spectra of these fluctuations 
show consistently the characteristic imprint of the two-dimensional theory of turbulence, 
which predicts a broken-power-law spectrum with exponents $p_1 \sim 5/3$ and $p_2 \sim 3$
on very general grounds (e.g., \cite{kraichnan,boffetta}). 
In this section we thus perform a 
systematic statistical study of turbulence spectra along the 768 different directions of our reference SN model, which 
will also allow us to compare
our findings with other assumptions most commonly found in the previous literature.

In order to compare the power spectra of the matter density profiles along different
directions, it is useful to fix a common range $\Delta r$ over which the Fourier transforms will be evaluated.
As commented in the context of Fig.~2, we assume $\Delta r = 2.1 \times 10^4$ km, 
corresponding to the distance between the radii of the mean forward and reverse shocks.
This assumption is not crucial for our findings (as we have explicitly checked), 
and is adopted hereafter only for the sake of
clarity and of homogeneity in the comparison of spectra at different angles.
With this in mind, for each direction $i$, we evaluated the power spectrum
over the range $[r_F^i - \Delta r ,r_F^i]$, in which $r_F^i$ is the radius of the forward
shock front along the $i$-th direction (solid curve in Fig.~2).

In   the region where matter
effects are relevant [see Eq.~(\ref{eq:res})], the neutrino oscillation length in matter is 
approximately given by~\cite{Fogli:2006xy}
\begin{equation}
\label{hierarchy}
\lambda_m \simeq 60 \,\ \textrm{km}\,\ 
\left(\frac{E}{10 \,\ \textrm{MeV}}\right)
\left(\frac{2\times 10^{-3}\,\ \textrm{eV}^2}{\Delta m^2_{\rm atm}} \right) \,
\left(\frac{0.2}{\sin 2 \theta_{13}}\right) \ , 
\end{equation}
where in our numerical calculations
 we take a reference value $\sin^2 \theta_{13} = 0.02$, consistent
with recent global data analyses \cite{Fogli:2012ua} (and corresponding to $\sin2\theta_{13}= 0.28$).
Therefore, it turns out that the radial simulation resolution $\delta r$, the neutrino wavelength in matter $\lambda_m$, and the 
radial range $\Delta r$ chosen for the Fourier transform, satisfy the correct length-scale hierarchy expected for 
a realistic representation of small-scale fluctuations,
\begin{equation}
\delta r < \lambda_m \ll \Delta r \,\ ,
\end{equation}
in the whole energy range relevant for SN neutrinos. 

This hierarchy suggests that the radial resolution of our reference SN simulation is good enough to characterize
the small scales relevant for neutrino oscillations in a turbulent environment, and (secondarily) that the shock wave extends
over many oscillation cycles.  Therefore, it makes sense to derive the spectral properties of the matter fluctuations directly from the matter density profiles, taken at face value along  different
directions. We anticipate, however, that the hierarchy in Eq.~(\ref{hierarchy}) is not necessarily guaranteed at large radii
by the worsening of the transverse resolution; this issue will be separately discussed in Sec.~5.

For the spectral analysis, we define the density contrast
\begin{equation}
\Delta\rho_i(r)=\frac{\rho_i(r)-\bar\rho_i}{\bar\rho_i}\ ,
\end{equation}
where $\bar\rho_i$ is the average density over the range $[r_F^i - \Delta r ,r_F^i]$, 
and expand it in a Fourier series:
\begin{equation}
\Delta\rho_i(r)=\sum_{n=-\infty}^{+\infty}c_i(n) e^{ik_n r}\ , \qquad \textrm{with} \qquad  k_n=\frac{2\pi n}{\Delta r}\ .
\label{eq:constrast}
\end{equation}
The Fourier transform
\begin{equation}
c_i(n) = \frac{1}{\Delta r} \int_{r_F^i - \Delta r}^{r_F^i} \Delta\rho_i(r) e^{-ik_n r } dr
\end{equation}
is then squared to give the power spectrum:
\begin{equation}
P_i(n) = 4 |c_i(n)|^2 \ .
\label{eq:power}
\end{equation}
In the following, we shall refer only to wave-number indices 
$n\geq 1$, since $c_i(-n)=c^*_i(n)$.

\begin{figure}[!t]
\centering
\includegraphics[width=0.325\columnwidth]{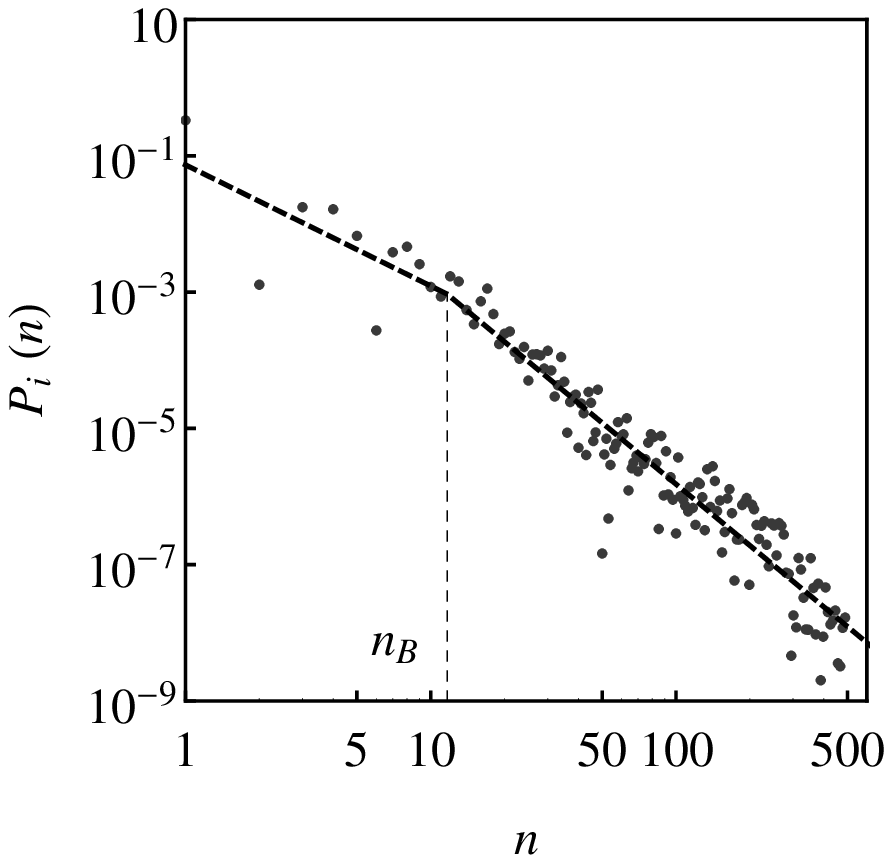}
\includegraphics[width=0.325\columnwidth]{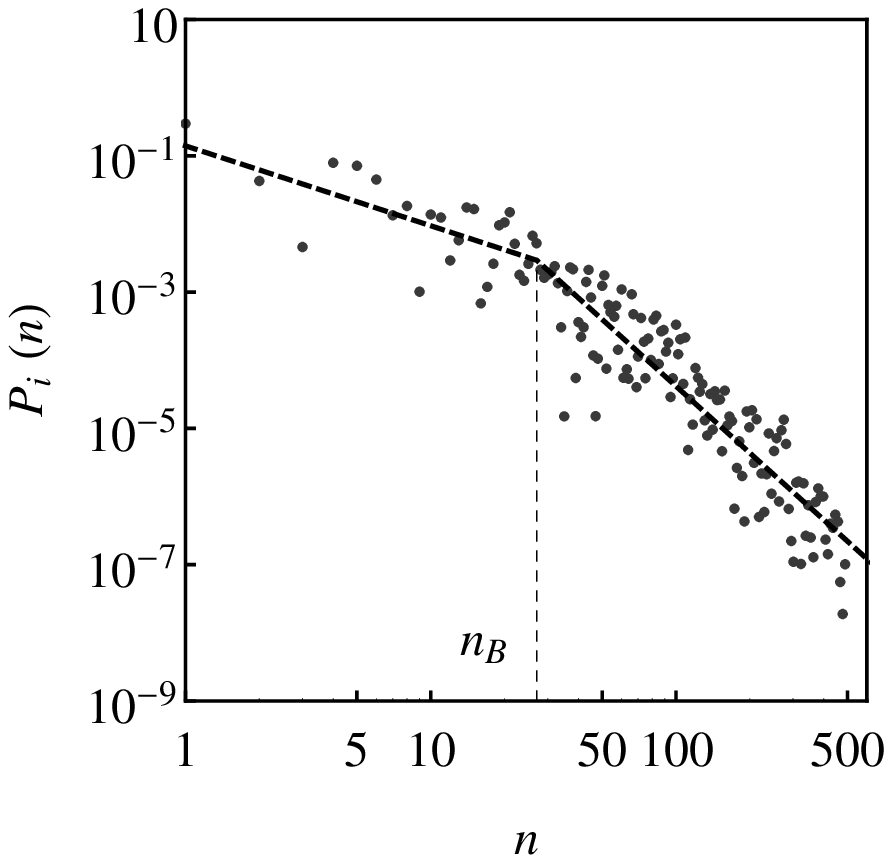}
\includegraphics[width=0.325\columnwidth]{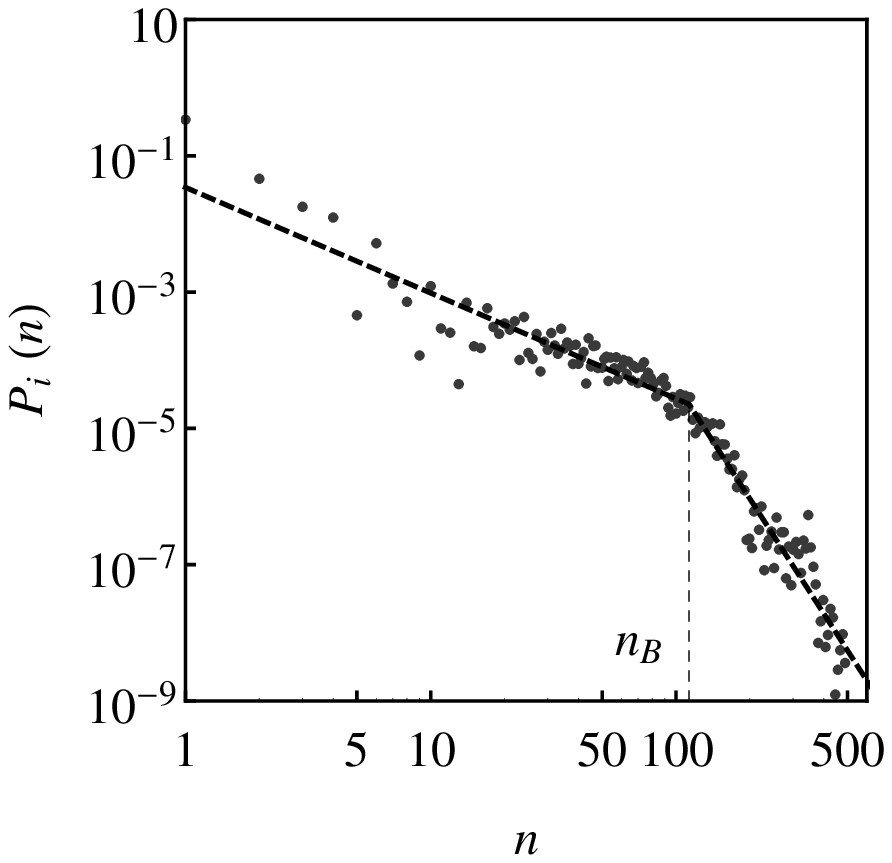}
\caption{Power spectra of the three matter density profiles 
in Fig.~3. They are representative of three different cases in which the break of the power spectrum appears 
at low (left panel, $n_B\sim10$), 
intermediate (central panel, $n_B\sim30$), 
and high (right panel, $n_B\sim100$) wave-number index.
} 
\label{fig:pow}
\end{figure}

Figure~\ref{fig:pow} shows the $P_i(n)$ values (dots) as a function of $n$ for 
the same three directions  of Fig.~\ref{fig3}
(corresponding to $i=9,\,177,\,466$). The points appear to be clustered along a rapidly 
decreasing curve, which, however, cannot be represented by a single slope in any of these 
three cases. Indeed, in such cases (as well as in all other directions, not shown), 
we find that the Fourier spectrum is quite reasonably 
represented by a two-slope line, i.e.\ by a broken power-law of the kind
\begin{equation}
P_i(n)=\left\{\begin{array}{ll}
A^2_i\, n^{-p_1}              &\  \textrm{for }  n \leq n_B\ ,   \\ 
A^2_i\, n_B^{p_2-p_1}\,n^{-p_2}\     &\  \textrm{for }  n \geq n_B\ ,
\end{array}\right. \ 
\label{eq:fit}
\end{equation}
where $A$ is the overall amplitude of the power-spectrum and $n_B$ is the 
wave-number index where the power-law spectrum
changes the exponent from $-p_1$ to $-p_2$.

\begin{figure}
\centering
\includegraphics[width=0.36\columnwidth]{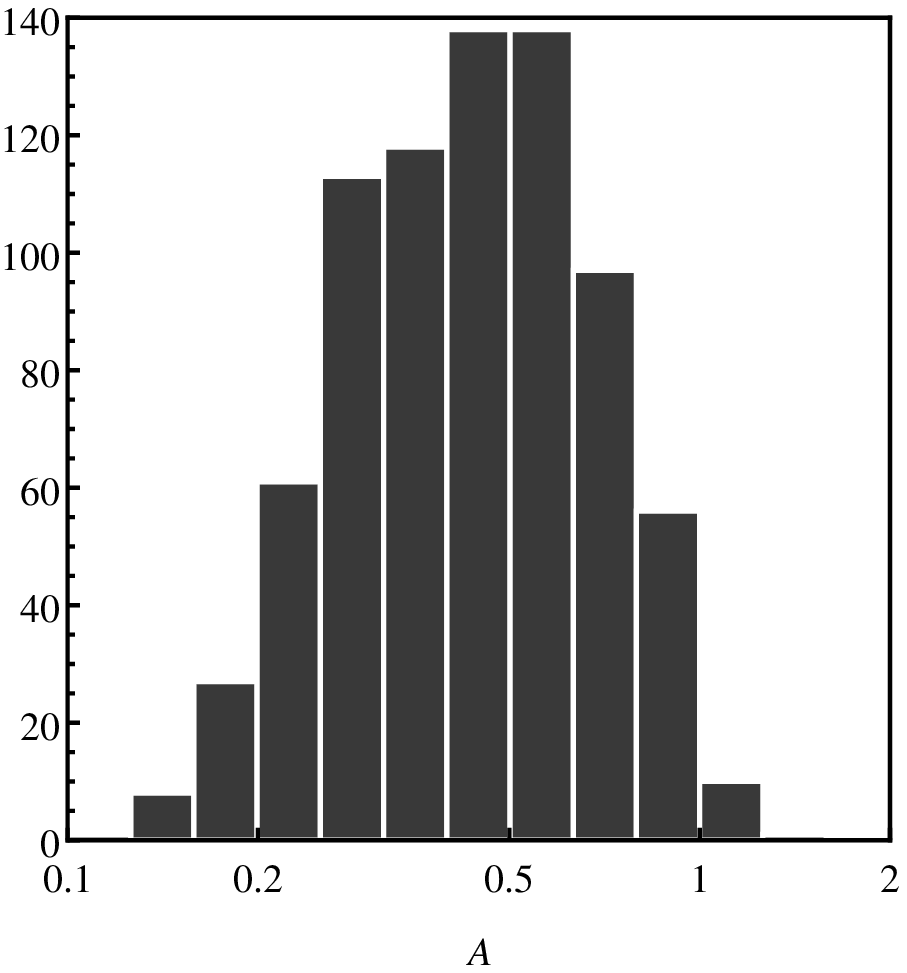}
\includegraphics[width=0.37\columnwidth]{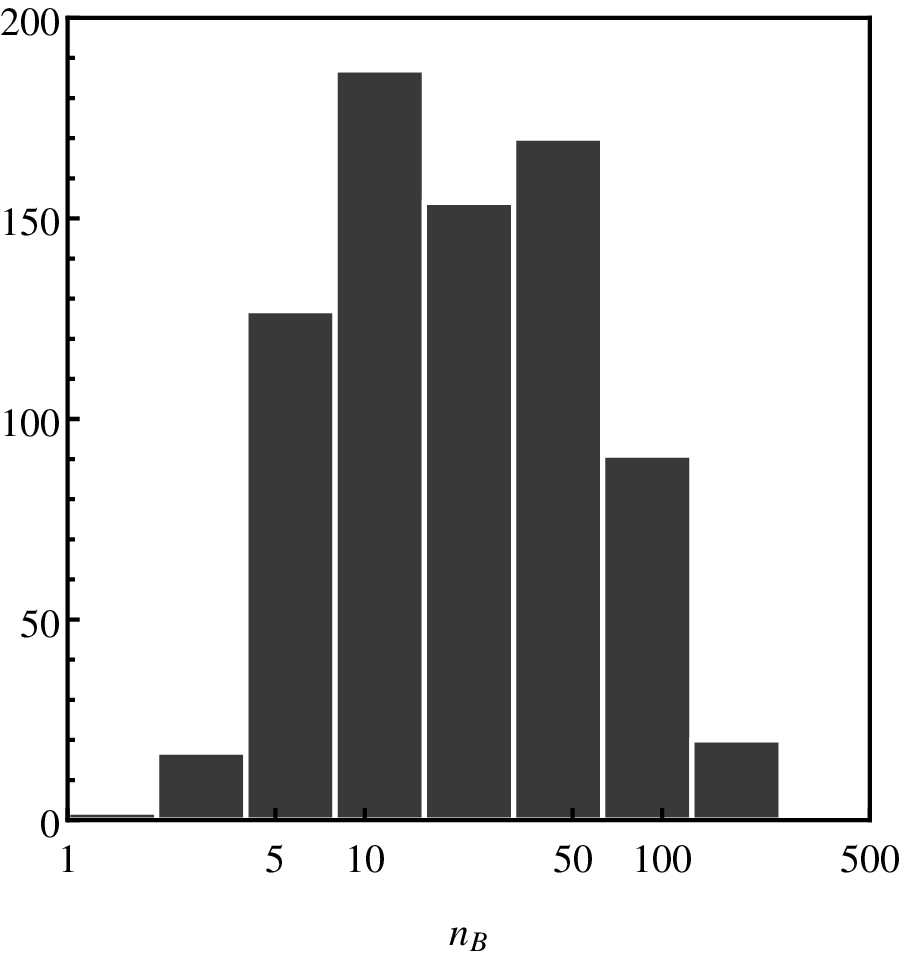}\\
\includegraphics[width=0.36\columnwidth]{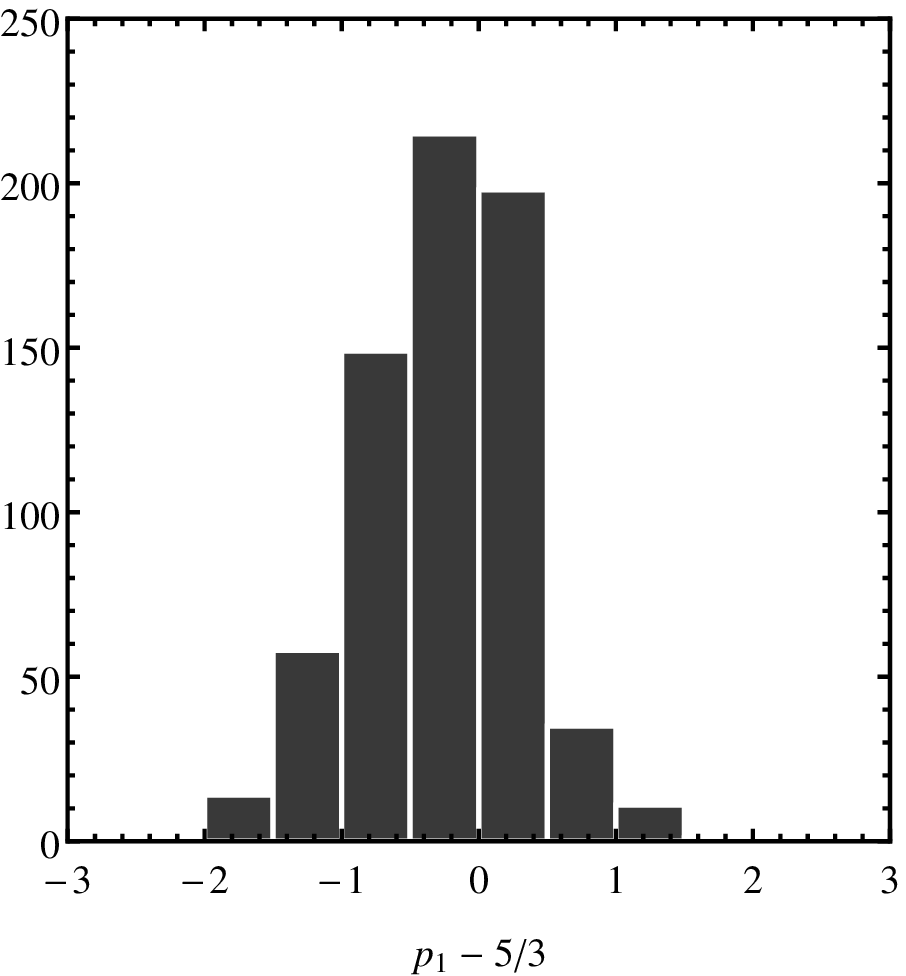}
\includegraphics[width=0.36\columnwidth]{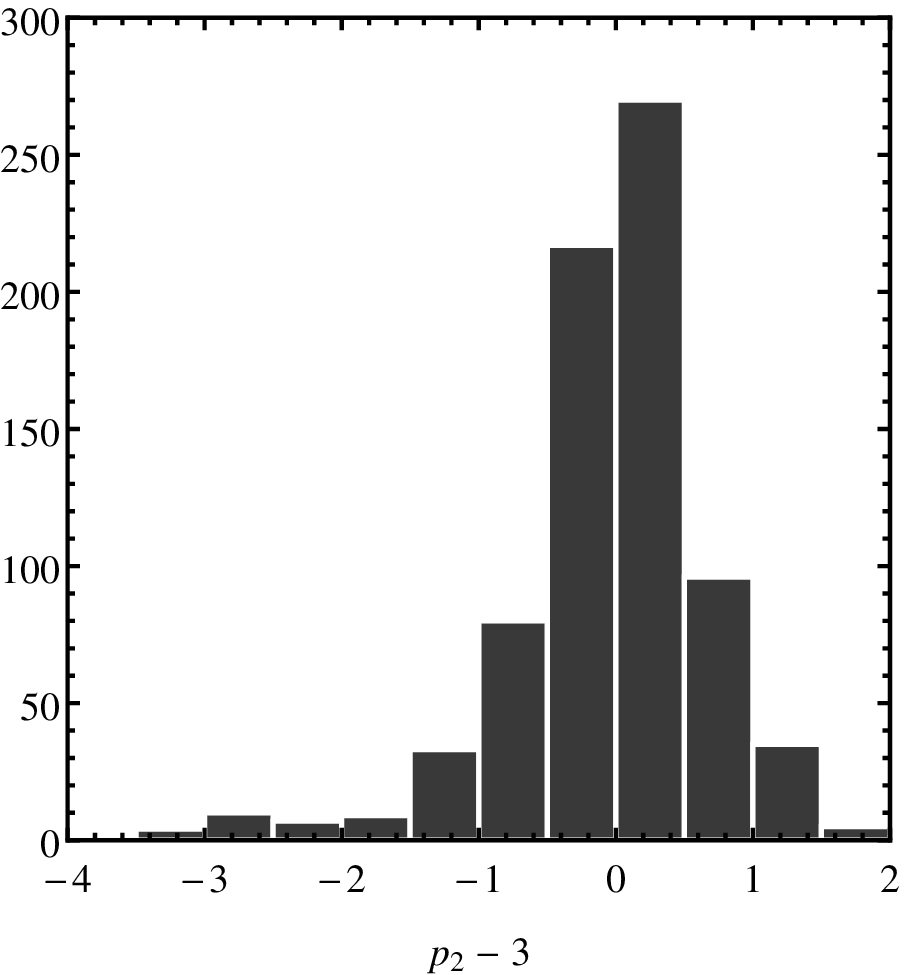}
\caption{Distributions of the four parameters $A$, $n_B$, $p_1$, $p_2$ of the 
broken-power law spectra of the density contrast [Eq.~(\ref{eq:fit})] 
for the $N=768$ radial directions.}
\label{fig:hist}
\end{figure}

For each of the 768 directions, we fit $P_i (n)$ in terms of four independent parameters $A$, $n_B$, $p_1$, and $p_2$ as in (\ref{eq:fit}), 
see dashed thick lines in Fig.~\ref{fig:pow}.
The distributions of the fit parameters are shown in Fig.~\ref{fig:hist},
where either linear or logarithmic scale was chosen in order to obtain more symmetric and ``gaussian''
distributions.
 The peak values and  1$\sigma$ range of the fit parameters are reported here:
 \begin{eqnarray}
 A &=& 0.44^{+0.24}_{-0.17}\ , \nonumber \\
n_B &=& 21^{+38}_{-14}\ , \nonumber \\
p_1 - 5/3 &=& 0.18^{+0.54}_{-0.77}\ , \nonumber \\     
p_2 - 3 &=& -0.04^{+0.57}_{-0.63}\ .
\label{eq:bestfit}
 \end{eqnarray}

The spectral indices, $p_1$ and $p_2$, of the two slopes of the broken power law are already 
expressed in terms of the values predicted for them by the 
Kolmogorov-Kraichnan theory
of two-dimensional turbulence~\cite{kraichnan}, namely $5/3$ and $3$,
respectively (see~\cite{boffetta} for a recent review),
which are indeed compatible with our results at $1\,\sigma$.

\begin{figure}
\centering
\includegraphics[width=0.5\columnwidth]{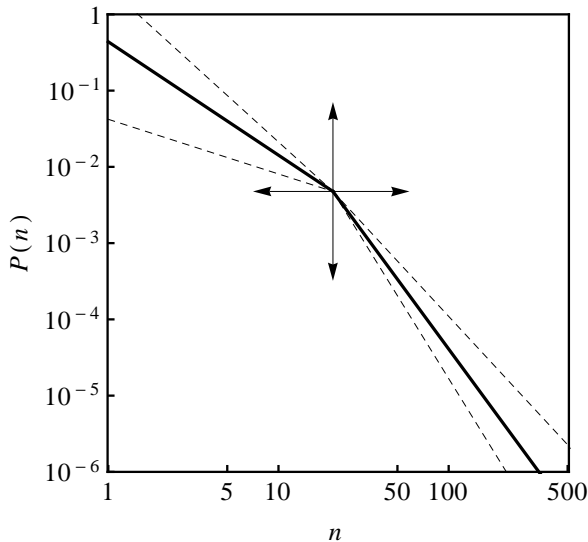}
\caption{\label{fig:synopsis} Synopsis of fit results 
in terms of a broken power law (thick line), with exponent changing from $p_1$ to $p_2$ at some wave number $n_B$. The 
$\pm 1\sigma$ errors on $n_B$ and on $P(n_B)$ are represented by a cross, while the $\pm1\sigma$ uncertainties of the exponents $p_1$ and $p_2$ are represented by dashed lines with different slopes. The power spectrum is compatible with the expectations
from the Kolmogorov-Kraichnan theory of 2D turbulence; see the text for details. 
 }
\end{figure}

Finally, Fig.~\ref{fig:synopsis} shows the synopsis of our fit results in terms of a broken power law with spectral indices $p_1$ and $p_2$ as in Eq.~(\ref{eq:fit}) (thick line) and 1\,$\sigma$
error ranges (dashed lines). The possible horizontal and vertical shift of the break is also shown, and takes into account the uncertainty at 1$\,\sigma$ C.L. on the determination of $n_B$ and of $P(n_B)$.
Our findings show that 2D turbulence imprints clearly emerge with the expected spectral features
from our reference SN simulation. We have found similar spectral results also for other post-bounce times (not shown).

\section{Neutrino crossing probabilities} 

In the previous section
we have shown that the reference SN density profiles embed the expected turbulence patterns, and trace
them with a space resolution appropriate to study flavor transitions in matter. We now analyze such transitions
in detail.

The main theoretical quantity associated to neutrino flavor change in SNe is the so-called crossing
 probability $P_H$, taking the same form for neutrinos and antineutrinos~(see \cite{Fogli:2003dw} and refs.\ therein)
\begin{equation}
P_H = P_H(\Delta m^2_{\rm atm},\, \sin^2 \theta_{13},\, E,\,\rho_i(r)) \,\ .
\end{equation} 
The crossing probability parameterizes the deviations with respect to a pure adiabatic
neutrino flavor evolution
in matter~\cite{Kuo:1989qe}. Indeed, violations of the adiabaticity 
would imply possible 
level crossings among the instantaneous neutrino mass eigenstates in matter. 
These violations could be  relevant especially at the resonance points [Eq.~(\ref{eq:res})].
For a non-monotonic matter density profile like the one encountered by neutrinos in SN, 
multiple level crossing can occur. In this case, one expects that the strongest violation
of adiabaticity would occur  when neutrinos propagate across the shock-front discontinuity. 
In this case, the crossing probability $P_H \simeq 1$. Conversely a purely adiabatic propagation would correspond to
$P_H=0$. The survival probability $P_{ee}$ of SN neutrinos at Earth (neglecting Earth matter crossing) is related 
to the crossing probability $P_H$ as follows: 
\begin{equation}
\label{cases} P_{ee} \simeq  \left\{
\begin{array}{ll}
 \sin^2\theta_{12}\, P_H & (\nu,\;\mathrm{NH}), \\
 \cos^2\theta_{12}       & (\overline\nu,\;\mathrm{NH}), \\
 \sin^2\theta_{12}       & (\nu,\;\mathrm{IH}), \\
 \cos^2\theta_{12}\, P_H & (\overline\nu,\;\mathrm{IH}). \\
\end{array}\right.
\end{equation}
From the above equations, it appears that $P_H$ can modulate the (otherwise constant) survival probability  
of $\nu_e$ in normal hierarchy and of $\bar\nu_e$ in inverted hierarchy, thus providing an important
handle to solve the current hierarchy ambiguity.

In order to compute $P_H$, we have numerically integrated flavor evolution equations along all the available matter 
density profiles, although we shall focus only on the representative ones shown in Fig.~3, for the sake
of simplicity.
We closely follow the approach of Ref.~\cite{Fogli:2003dw} to compute $P_H$ 
(the interested reader is referred to this reference for further details). In particular,
since matter effects are dominant at large radii
($r \gtrsim 10^3$~km) we neglect  non-linear effects associated with neutrino-neutrino interactions that  occur
at smaller radii.

 \begin{figure}[t]
 \centering
\includegraphics[width=0.31\columnwidth]{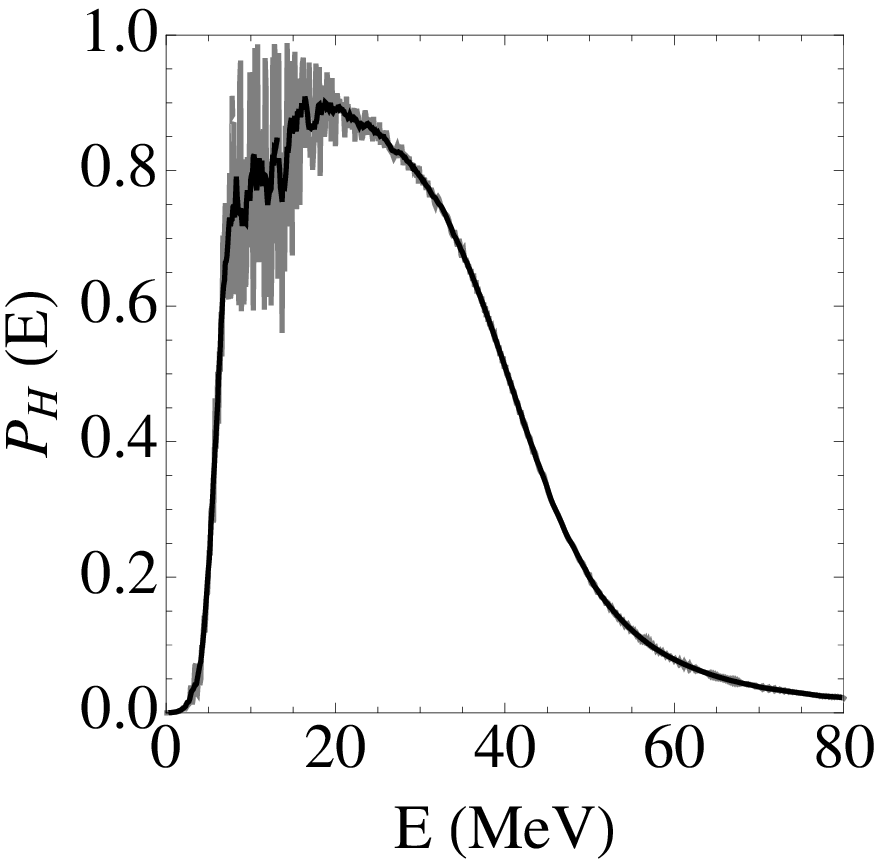}
 \includegraphics[width=0.31\columnwidth]{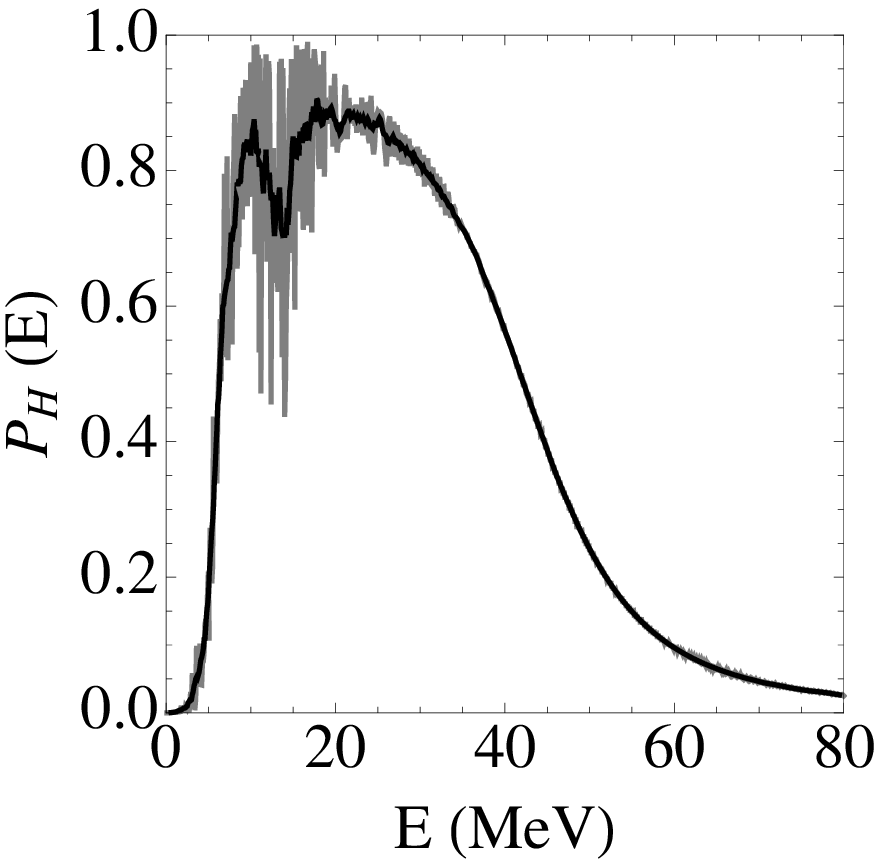}
 \includegraphics[width=0.31\columnwidth]{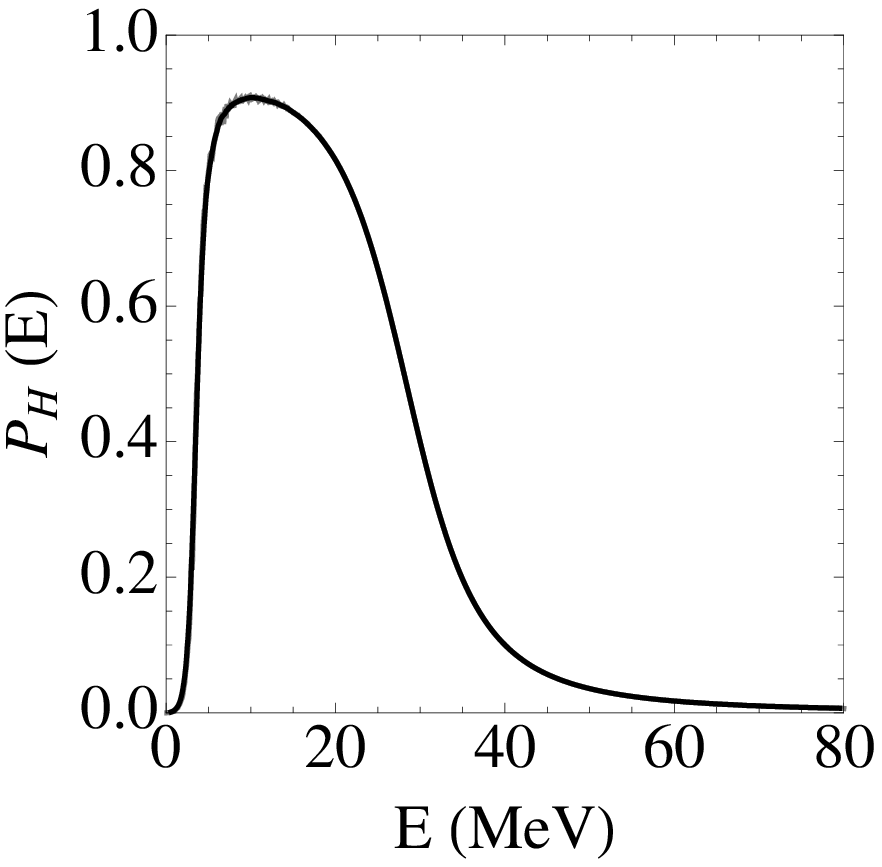}
 \vspace*{-2mm}
 \caption{Neutrino crossing probability $P_H$ as function of energy $E$ for the three SN matter density profiles of Fig.~3
(gray curves). In the black curves, phase effects are averaged out.  
}  \label{fig7}
 \end{figure}

In Fig.~7 we depict the crossing probability $P_H$ for the three chosen representative cases, as a 
function of the neutrino energy $E$ (gray curves). 
In all cases, $P_H$ exhibits the  typical top-hat structure widely discussed in previous literature (as quoted, e.g., in~\cite{Fogli:2003dw}). In particular, $P_H$  rapidly jumps from zero  to $\sim 1$ for $E\sim 10$~MeV. 
  The energy range where   $P_H \sim 1$ corresponds to resonances
 occurring within  the forward shock front, 
where extremely non-adiabatic transitions take place. 
 Indeed, as shown in~\cite{Fogli:2003dw}, for values of $\sin^2 \theta_{13}$ as large
as the measured one, the crossing probabilities occurring before the shock front and along the static
profile ahead of the shock are exponentially suppressed.
Instead, the energy width of the top-hat function depends
 on the height of the shock front relative to the resonance band in 
 Fig.~3. In the left and central panels  the resonance band intersects the forward shock 
front for all the range $[5,50]$~MeV. Therefore, in all this energy range  the resonance occurring
within 
the shock front gives $P_H \sim 1$, while the value of $P_H$ 
drops to zero outside this energy range. 
Conversely, in the right panel the energy range of $P_H \sim 1$ is smaller,
since   the resonance occurs on the static profile
already at $E\gtrsim 40$~MeV, where $P_H$
drops to zero.

We also realize that, while in the right panel the $P_H(E)$ function has a smooth profile, in the other panels it presents an oscillatory
behavior in the energy range around $E\in [10,30]$~MeV.   This oscillatory behavior is due to the phase effect caused by the 
interference between  multiple narrowly-spaced resonances~\cite{Dasgupta:2005wn} corresponding to a rapidly and largely 
fluctuating turbulent profile (see Fig.~3). In practice, the observability of this oscillatory pattern 
is  severely limited by the inability of the detectors to reconstruct the neutrino energy
precisely~\cite{Dasgupta:2005wn}, leading to a smeared $P_H$ profile. 
For illustration, the black curves in Fig.~7 are obtained by convoluting the gray ones 
with a ``top-hat'' energy resolution function having $\pm 0.5$~MeV width, so as to enforce an effective
phase averaging (smearing).

Independently of the details, we find that $P_H$ can be sizable in all the three panels
of Fig.~7, and is not significantly suppressed in those cases where fluctuations are evident (left and middle
panels) as compared with the case having a smoother matter profile (right panel). Moreover, although the three 
corresponding matter density profiles are associated with power spectra having a break at rather
different $n_B$ (see Fig.~4), the global behavior of $P_H$ is rather similar, i.e., 
 there is no evident link between the break in the power spectrum and the behavior
 of $P_H$.
  Indeed, what is mostly relevant for the shape of  $P_H$ is the presence of a crossing (resonance) within
the forward shock. 
 This fact is   mostly related to the height of the shock front rather than to the presence of the turbulence.
 Indeed, as already noticed in~\cite{Lund:2013uta}, the ``large'' measured value of $\theta_{13}$ 
makes the resonances more adiabatic downstream of the forward shock front. Then, the main effect of turbulence
is the phase effect which increases with the amplitude of the fluctuations.  In addition, 
 in none of the cases considered the turbulence has enough strength
to strongly ``damp'' the profile of the $P_H$.
Large damping (with $P_H \sim 0.5$) was found in~\cite{Fogli:2006xy} where ``delta-correlated'' fluctuations were considered, with
correlation scale of $L=10$~km and a fluctuation amplitude  of a few $\%$. Such a hypothetical amplitude is actually 
much larger than what our simulation
gives on the same length scale. 
Indeed,  the density contrast on a scale $L \sim 10$~km, is  given by $\Delta \rho \sim A (\Delta r/L)^{-p_1/2} \sim 
3 \times 10^{-4}$ for our best fit values [see Eq.~(\ref{eq:bestfit})].

In~\cite{Lund:2013uta}  a Kolmogorov-like fluctuation power-spectrum 
(characterized by a single exponent $p_1=5/3$ in 3D turbulence theory) was included 
through the 
 ``randomization method''~\cite{random}
 on top of an undisturbed density profile. A parametric study was performed for different values of 
the overall amplitude of the fluctuation power-spectrum, including cases with ``small'' values (of the same order of those emerging
in our reference SN model). In particular, for a  power-spectrum amplitude $A$ of $0.1-0.3$, 
comparable with the  amplitude $A$ from our simulations (see Fig.~3),
no significant damping was found, while 
phase effects were still present. We qualitatively agree with such findings, but the intrinsic differences between our 
SN models (plus the fact that collective and matter effects are combined in \cite{Lund:2013uta}) prevent 
a more detailed comparison. In any case, we remark that our results are directly linked to a simulation  
rather than to a parameterization.

\section{Reconstructing small scale fluctuations with the randomization method}

In the previous Section the crossing probability ${P_H}$ has been evaluated by integrating
the neutrino flavor evolution equations along same representative density profiles, taken directly
from our reference 2D simulations. As discussed in Sec.~2, for the time snapshot we are considering
($t=2$~s)
these simulations have the best \emph{radial} resolution $(\delta r=15.3$~km). However, given the angular
separation of $\Delta \theta \sim 0.234^\circ$ among two radial consecutive trajectories, it follows that
the \emph{transverse} resolution  is given by
\begin{equation}
{\delta r}_\perp = 4.09 \,\ \textrm{km} \left(\frac{r}{10^3 \,\ \textrm{km}}\right) \,\ .
\end{equation}
Therefore, at the relatively large radii characterizing the shock front 
(average value $r_F=2.7 \times 10^4$~km), where flavor conversions largely develop, 
the transverse resolution can be  
${\delta r}_\perp \simeq 110$~km, which is definitely worse than the radial one.
 
Since turbulence develops both in radial and transverse directions, this fact implies that the simulations
might not really resolve length scales below 
${\mathcal O(10^2)}$~km at typical shock-front radii. From the viewpoint of the power spectrum $P(n)$, this 
length scale corresponds to a multipoles number $n_\perp \simeq 200$. In other words, the spectrum features
for $n>n_\perp$ might not be correctly captured by the available simulations.

We stress that this limitation does not invalidate the statistical analysis of the power spectra in Sec.~3. Indeed,
in the vast majority of analyzed cases the spectrum ``knee'' (i.e., the multipole $n_B$) is found below $n_\perp \simeq 200$
(see the upper right panel of Fig.~5). However, an overall resolution of O(100)~km can be of the same order or larger
than typical oscillation lengths, and may invalidate the hierarchy in Eq.~(\ref{hierarchy}), with possible alterations 
of the $P_H(E)$ profiles discussed in Sec.~4.

In order to overcome this limitation and to investigate the effect of small scales not directly 
resolved by the simulations,
it is necessary to rely on some procedure to ``continue'' the power-spectrum for multipoles larger than $n_\perp$.
At this regard, in the literature it has been proposed to generate random fluctuations with a given
power spectrum by means of the 
``randomization method''~\cite{random}. In~\cite{Kneller:2010sc,Kneller:2013ska,Lund:2013uta}
this method was applied to generate  fluctuations over smooth 1D density profiles, assuming 
a Kolmogorov spectrum with tunable amplitudes.
Herein, we investigate  a variant of this randomization procedure. We take the previous (Sec.~3) fluctuation spectra
at face value for $n\leq n_\perp$, and randomize them only for $n>n_\perp$; for definiteness we take $n_\perp=200$, although
the precise number chosen for $n_\perp$ is of little relevance. We also anchor the small-scale randomization to large-scale 
spectral features as follows.

After truncating the power spectra at $n_\perp$, we have studied the distribution of the amplitudes $|c_n|= \frac{1}{2}\sqrt{P(n)}$ 
[see Eq.~(\ref{eq:power})] along each direction. In particular, we have considered their residuals with respect to the
values $|{\bar c}_n|$ expected from the best-fit, broken power-law spectrum (as represented, e.g., by the continuous lines
in Fig.~4). As it can be seen at a glance in Fig.~4 (as well as for other profiles, not shown), the typical scatter of
residuals is relatively similar in different profiles (in logarithmic scale). We found that the distribution of the fractional 
log-residuals, defined as
\begin{equation}
\delta_n = \frac{\log |c_n|-\log |{\bar c}_n|}{\log |{\bar c}_n|} \,\ .
\label{eq:resid}
\end{equation}
is approximately gaussian in each of the 768 truncated spectra, with central value always close to zero, and 
variance quite similar for all the trajectories, within a factor less than two. 

A conservative estimate for the overall variance of the residuals is  $\sigma^2 \simeq (0.02)^2$, for $n\leq n_\perp$.   
We then assume that the $|c_n|$ values for $n>n_\perp$ are also scattered around the best-fit
Kraichnan-Kolmogorov spectrum
with this same overall variance, and with a random phase $\phi_n$. More precisely, for $n >  n_\perp$
we randomly generate amplitudes as
\begin{equation}
c_n= |{\bar c}_n|^{\delta_n+1}e^{i\phi_n} \,\ , 
\label{eq:genercn}
\end{equation}
where the $\delta_n$ are drawn from a gaussian distribution with null average and 
variance $\sigma^2 \simeq (0.02)^2$, and the phases $\phi_n$ are drawn from a flat distribution in $[0,2\pi]$.
Then the density contrast  in each direction [Eq.~(\ref{eq:constrast})] is derived as
\begin{equation}
\Delta \tilde{\rho}_i = \left(\sum_{n \leq n_\perp} + \sum_{n > n_\perp} \right) c_n e^{i k_n r} \,\ ,
\end{equation}
 where for   the coefficients $c_n$ are taken directly from the simulations ${n \leq n_\perp}$, while they are  
randomized  via Eq.~(\ref{eq:genercn}) for $n > n_\perp$. Of course, one can generate different realizations of 
the density profiles for $n > n_\perp$ (see below). Finally, 
we adopt a filter in order to suppress spurios localized 
fluctuations (spikes) in the reconstructed density profiles, which may emerge in singular points
where the original profile is step-like.

 \begin{figure}[!t]
 \centering
 \includegraphics[width=0.4\columnwidth]{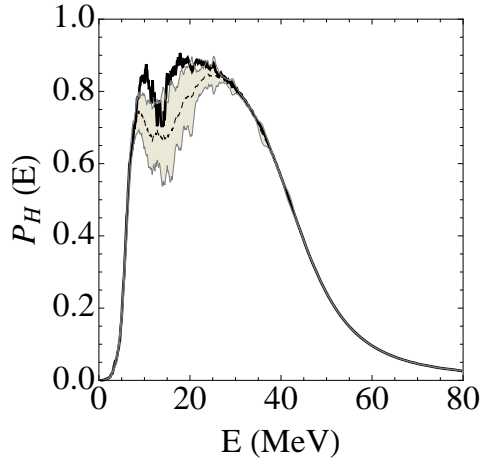}
 \caption{Phase-averaged crossing probability $P_H$: comparison of the results reported in the central panel of Fig.~7 (thick line)
 with the results obtained from 100 realizations of the randomization procedure (dashed line with $\pm1\sigma$ dispersion gray band).
} \label{fig8}
 \end{figure}

The density profile in the central panel of Fig.~3 is particularly well suited to investigate 
the effects of the above randomization procedure, since it displays large
fluctuations in the interesting energy region and for a large radial range. Conversely,
 in the left (right) panel of Fig.~3, level crossings occur only at small radii (energies), where 
 randomization effects produce only a minor impact on $P_H$ (not shown).
 
Focusing on the profile in the central panel of Fig.~3, we have calculated the crossing probability $P_H(E)$ 
for 100 realizations of the randomized density fluctuations above $n_\perp$. We apply energy smearing as in Fig.~7 
to remove (unobservable) phase effects in each realization. 
Figure~8 shows the average $P_H(E)$ profile from the 100 realizations (dashed curve), together with its $\pm 1\sigma$
dispersion (gray band). For comparison, the $P_H(E)$ profile in Fig.~7 is superposed (thick black curve). It appears
that in the energy range $E\in [20,40]$~MeV the effect of (randomized) small scale fluctuations 
can generate differences up to $\sim 20\%$ in the $P_H$, but does not radically change the profile. 
In particular, the double-peak shape of the $P_H$ is similar to the one obtained from the original density profile, 
taken at face value from the simulations. Moreover, the results from this profile
are roughly in agreement within the $\pm1\sigma$ dispersion band of the randomized profile. 
In general, by applying the randomization method described above also to other density profiles from
our 2D simulations,  we do not find dramatic changes (e.g., strong or complete depolarization) as in other cases
considered in the previous  literature~\cite{Fogli:2006xy,Kneller:2010sc,Kneller:2013ska,Lund:2013uta}.
Although limited to 2D SN models so far, our results show the importance of anchoring studies of turbulence effect and randomization procedures, whenever possible, to high-resolution SN simulations.


\section{Summary and prospects}
 
In this work we have carried out a study of the effects of matter density 
fluctuations on the conversion probabilities of SN neutrinos, characterizing the matter turbulence
from a high-resolution 2D supernova model. The latter was evolved to 
sufficiently late times to follow the shock dynamics
in the density regime relevant for flavor transformations at energies typical of supernova neutrinos.
We found that the power spectra of the matter density fluctuations along different radial directions 
show a  ``typical'' structure which represents the imprint of generic 2D turbulence, namely 
a broken power-law spectrum with average exponents $p_1 \sim 5/3$ and $p_2 \sim 3$. 
We evaluated the flavor evolution of SN neutrinos along representative density profiles, exhibiting 
power-spectra with breaks at different wave numbers. We found that the crossing probability $P_H$ exhibits,
in all the different cases, a top-hat structure where $P_H$ jumps to
$\sim 1 $ at energies where an extremely non-adiabatic resonance takes place across the shock front. 
We do not find significant damping of the crossing probabilities associated with the matter
turbulence. Since the available simulations might not fully resolve details at
length scales below O(100) km, we have also improved the treatment of small-scale fluctuations 
by means of a randomization procedure anchored to larger scales.  We find that the randomized profiles 
generally preserve the main features of the $P_H(E)$ profile, although with a slightly more pronounced damping
in some cases.
As a result, the shock-wave imprint on the crossing probability seems to be
a possible observable signature in the future core-collapse SN events, with no large suppression
by turbulence. We emphasize that, if SN neutrino energy spectra of different flavor are not too similar 
at late times~\cite{Fischer:2009af}, shock-wave effects would
give an observable signature in the SN neutrino signal at large underground detectors~\cite{Fogli:2004ff,Tomas:2004gr}. 
This would be an important tool 
to follow in real-time the SN explosion dynamics, as well as to determine the neutrino mass hierarchy.  
Therefore, we hope that the potential importance of shock-wave  signatures as 
a probe of neutrino physics and astrophysics  will motivate further attention and progress in numerical
simulations, in order to investigate the impact of matter turbulence in other independent SN models, 
possibly also in 3D. 

As already commented, the transition from 2D to 3D is expected to bring turbulence spectra from a broken power law 
into a single power law, with a
Kolmogorov exponent $p_1=5/3$ across all scales. This expectation (not yet confirmed by the available 3D 
results~\cite{Dolence:2013,Couch:2013b})
would in principle bring more power to fluctuations at small scales, with 
a possible enhancement of the damping effect. In order to clarify this issue, a detailed study 
with high-resolution three-dimensional SN models is mandatory. In this context, another delicate 
issue is represented by the poorly known
asymmetries in the progenitor stars prior to collapse. While all current SN simulations are
started from spherical models, it is possible that the progenitor core at the onset of
gravitational instability has turbulent density perturbations on different scales
(e.g., \cite{Arnett:2011} and references therein). These pre-collapse inhomogeneities and 
fluctuations could affect the SN explosion by seeding and
enhancing the hydrodynamic instabilities in the post-shock region~\cite{Couch:2013a},
and their effects on MSW neutrino flavor conversions still need to be explored.
Finally, it should be noted that, since the neutrinosphere source has a finite size, the neutrinos emitted from different points propagate through somewhat different (but correlated) density profiles. In this context, it might also be interesting to study 
the correlation of the transition probabilities along relatively close trajectories, as proposed in~\cite{Kneller:2013lka}.

In conclusion, we believe that our study offers an interesting new perspective to the issues raised by
SN neutrino turbulence, at least in the realm of 2D simulations. Although our approach has some limitations which prevent
a wide generalizatios of the results, especially to the 3D case, it illustrates a methodology that can be usefully applied also to other SN models.  When future, high-resolution simulations will become available for detailed and comparative analyses,  
further studies will test the robustness of our results.

\section*{Acknowledgements}

We warmly thank Konstantinos Kifonidis for providing the
data of his 2D simulations and Timur Rashba, Lab Saha, Irene Tamborra, and Ricard Tom{\`a}s for useful discussions 
and help on various topics related to this work. 
The work of E.B.\ and A.M.\ was supported by the German Science Foundation (DFG) within the Collaborative Research Center 676 ``Particles, Strings and the Early Universe''. The work of E.L.\ was partially supported by the Italian MIUR and 
Istituto Nazionale di Fisica Nucleare (INFN) through the ``Theoretical Astroparticle
Physics'' project. H.T.J.\ acknowledges support by the Deutsche Forschungsgemeinschaft through the
the Transregional Collaborative Research Center 
SFB/TR-7 ``Gravitational Wave Astronomy'' and the Cluster of Excellence EXC~153 ``Origin and Structure 
of the Universe.''
S.C.\ acknowledges support from the European Union through a Marie Curie Fellowship, Grant No.\ PIIF-GA-2011-299861, and through the ITN ``Invisibles'', Grant No.\ PITN-GA-2011-289442.



\begin{thebibliography}{99}

 \bibitem{Pantaleone:1992eq}  
  J.~Pantaleone,  
  ``Neutrino oscillations at high densities,''  
  Phys.\ Lett.\ B {\bf 287}, 128 (1992).  
  
\bibitem{Qian:1994wh}
  Y.~Z.~Qian and G.~M.~Fuller,
  ``Neutrino-neutrino scattering and matter enhanced neutrino flavor
  transformation in Supernovae,''
  Phys.\ Rev.\  D {\bf 51}, 1479 (1995)
  [astro-ph/9406073].
 
\bibitem{Duan:2005cp}
  H.~Duan, G.~M.~Fuller and Y.~Z.~Qian,
  ``Collective Neutrino Flavor Transformation In Supernovae,''
  Phys.\ Rev.\  D {\bf 74}, 123004 (2006)
  [astro-ph/0511275].
  

\bibitem{Duan:2006an}
  H.~Duan, G.~M.~Fuller, J.~Carlson and Y.~Z.~Qian,
  ``Simulation of coherent non-linear neutrino flavor transformation in the
  supernova environment. I: Correlated neutrino trajectories,''
  Phys.\ Rev.\  D {\bf 74}, 105014 (2006)
  [astro-ph/0606616].
  
\bibitem{Hannestad:2006nj}
  S.~Hannestad, G.~G.~Raffelt, G.~Sigl and Y.~Y.~Y.~Wong,
  ``Self-induced conversion in dense neutrino gases: Pendulum in flavour  
 space,''
  Phys.\ Rev.\  D {\bf 74}, 105010  (2006)
  [Erratum-ibid.\  D {\bf 76},  029901 (2007)]
  [astro-ph/0608695].
  
  

\bibitem{Fogli:2007bk}
  G.~L.~Fogli, E.~Lisi, A.~Marrone and A.~Mirizzi,
  ``Collective neutrino flavor transitions in supernovae and the role of
  trajectory averaging,''
  JCAP {\bf 0712}, 010 (2007)
  [arXiv:0707.1998 [hep-ph]].
  
  
  
\bibitem{Duan:2010bg} 
  H.~Duan, G.~M.~Fuller and Y.~-Z.~Qian,
  ``Collective Neutrino Oscillations,''
  Ann.\ Rev.\ Nucl.\ Part.\ Sci.\  {\bf 60}, 569 (2010)
  [arXiv:1001.2799 [hep-ph]].
 
\bibitem{Duan:2007bt} 
  H.~Duan, G.~M.~Fuller, J.~Carlson and Y.~-Q.~Zhong,
  ``Neutrino Mass Hierarchy and Stepwise Spectral Swapping of Supernova Neutrino Flavors,''
  Phys.\ Rev.\ Lett.\  {\bf 99}, 241802 (2007)
  [arXiv:0707.0290 [astro-ph]].
 
\bibitem{Dasgupta:2009mg} 
  B.~Dasgupta, A.~Dighe, G.~G.~Raffelt and A.~Y.~.Smirnov,
  ``Multiple Spectral Splits of Supernova Neutrinos,''
  Phys.\ Rev.\ Lett.\  {\bf 103}, 051105 (2009)
  [arXiv:0904.3542 [hep-ph]].
  
  
\bibitem{Mirizzi:2013wda} 
  A.~Mirizzi,
  ``Self-induced spectral splits with multi-azimuthal-angle effects for different supernova neutrino fluxes,''
  arXiv:1308.5255 [hep-ph].

\bibitem{Matt}  L.~Wolfenstein,  
				``Neutrino Oscillations In Matter,''  
                Phys.\ Rev.\ D {\bf 17}, 2369 (1978);  
                S. P.~Mikheev and A. Yu.\ Smirnov,  
                ``Resonance Enhancement Of Oscillations In Matter And Solar Neutrino  
				Spectroscopy,''  
                Yad.\ Fiz.\ {\bf 42}, 1441 (1985)  
                [Sov.\ J.\ Nucl.\ Phys.\ {\bf 42}, 913 (1985)].  
                
                
\bibitem{Dighe:1999bi}
  A.~S.~Dighe and A.~Y.~Smirnov,
  ``Identifying the neutrino mass spectrum from the neutrino burst from a
  supernova,''
  Phys.\ Rev.\  D {\bf 62}, 033007 (2000)
  [hep-ph/9907423].
  
\bibitem{Schirato:2002tg} 
  R.~C.~Schirato and G.~M.~Fuller,
  ``Connection between supernova shocks, flavor transformation, and the neutrino signal,''
  astro-ph/0205390.
  
\bibitem{Fogli:2003dw} 
  G.~L.~Fogli, E.~Lisi, D.~Montanino and A.~Mirizzi,
  ``Analysis of energy and time dependence of supernova shock effects on neutrino crossing probabilities,''
  Phys.\ Rev.\ D {\bf 68}, 033005 (2003)
  [hep-ph/0304056].
  
  \bibitem{Fogli:2004ff} 
  G.~L.~Fogli, E.~Lisi, A.~Mirizzi and D.~Montanino,
  ``Probing supernova shock waves and neutrino flavor transitions in next-generation water-Cerenkov detectors,''
  JCAP {\bf 0504}, 002 (2005)
  [hep-ph/0412046].
  
\bibitem{Tomas:2004gr} 
  R.~Tom{\`a}s, M.~Kachelriess, G.~Raffelt, A.~Dighe, H.~-T.~Janka and L.~Scheck,
  ``Neutrino signatures of supernova shock and reverse shock propagation,''
  JCAP {\bf 0409}, 015 (2004)
  [astro-ph/0407132].
  
\bibitem{Gava:2009pj} 
  J.~Gava, J.~Kneller, C.~Volpe and G.~C.~McLaughlin,
  ``A Dynamical collective calculation of supernova neutrino signals,''
  Phys.\ Rev.\ Lett.\  {\bf 103}, 071101 (2009)
  [arXiv:0902.0317 [hep-ph]].


\bibitem{Kifonidis:2003} 
K.~Kifonidis, T.~Plewa, H.-T.~Janka and E.~M\"uller,
  ``Non-spherical core collapse supernovae. I. Neutrino-driven convection, Rayleigh-Taylor instabilities, and the formation and propagation of metal clumps,''
  Astron.\ Astrophys.\  {\bf 408}, 621 (2003)
  [astro-ph/0302239].


\bibitem{Kifonidis:2006}  
K.~Kifonidis, T.~Plewa, L.~Scheck, H.-T.~Janka and E.~M\"uller,
  ``Non-spherical core collapse supernovae. II. The late-time evolution of globally anisotropic neutrino-driven explosions and their implications for SN 1987 A,''
  Astron.\ Astrophys.\  {\bf 453}, 661 (2006)
  [astro-ph/0511369].

\bibitem{Scheck:2006}
L.~Scheck, K.~Kifonidis, H.-T.~Janka and E.~M\"uller,
  ``Multidimensional supernova simulations with approximative neutrino transport. I. Neutron star kicks and the anisotropy of neutrino-driven explosions in two spatial dimensions,''
  Astron.\ Astrophys.\  {\bf 457}, 963 (2006)
  [astro-ph/0601302].

\bibitem{Hammer:2010}
N.~Hammer, H.-T.~Janka, and E.~M\"uller,
  ``Three-dimensional simulations of mixing instabilities in supernova explosions,''
  Astrophys.\ J.\  {\bf 714}, 1371 (2010)
  [astro-ph/0908.3474].

\bibitem{Arcones:2011}
A.~Arcones and H.-T.~Janka,
  ``Nucleosynthesis-relevant conditions in neutrino-driven supernova outflows. II. The reverse shock in two-dimensional simulations,''
  Astron.\ Astrophys.\  {\bf 526}, 160 (2011)
  [astro-ph/1008.0882].

\bibitem{Mueller:2012}
E.~M\"uller, H.-T.~Janka and A.~Wongwathanarat, 
  ``Parametrized 3D models of neutrino-driven supernova explosions. Neutrino emission asymmetries and gravitational-wave signals,''
  Astron.\ Astrophys.\  {\bf 537}, 63 (2012)
  [astro-ph/1106.6301].

\bibitem{Wongwathanarat:2013}
A.~Wongwathanarat, H.-T.~Janka, and E.~M\"uller,
  ``Three-dimensional neutrino-driven supernovae: Neutron star kicks, spins, and asymmetric ejection of nucleosynthesis products,''
  Astron.\ Astrophys.\  {\bf 552}, 126 (2013)
  [astro-ph/1210.8148].



\bibitem{Schaefer:1987fr} 
  A.~Schaefer and S.~E.~Koonin,
  ``Influence of Density Fluctuations on Solar Neutrino Conversion,''
  Phys.\ Lett.\ B {\bf 185}, 417 (1987).
  
\bibitem{Sawyer:1990tw} 
  R.~F.~Sawyer,
  ``Neutrino oscillations in inhomogeneous matter,''
  Phys.\ Rev.\ D {\bf 42}, 3908 (1990).
  
\bibitem{Loreti:1994ry} 
  F.~N.~Loreti and A.~B.~Balantekin,
  ``Neutrino oscillations in noisy media,''
  Phys.\ Rev.\ D {\bf 50}, 4762 (1994)
  [nucl-th/9406003].
  
\bibitem{Nunokawa:1996qu} 
  H.~Nunokawa, A.~Rossi, V.~B.~Semikoz and J.~W.~F.~Valle,
  ``The Effect of random matter density perturbations on the MSW solution to the solar neutrino problem,''
  Nucl.\ Phys.\ B {\bf 472}, 495 (1996)
  [hep-ph/9602307].
  
\bibitem{Balantekin:1996pp} 
  A.~B.~Balantekin, J.~M.~Fetter and F.~N.~Loreti,
  ``The MSW effect in a fluctuating matter density,''
  Phys.\ Rev.\ D {\bf 54}, 3941 (1996)
  [astro-ph/9604061].

\bibitem{Burgess:1996mz} 
  C.~P.~Burgess and D.~Michaud,
  ``Neutrino propagation in a fluctuating sun,''
  Annals Phys.\  {\bf 256}, 1 (1997)
  [hep-ph/9606295].
  
  \bibitem{torrente}
 E.~Torrente-Lujan,
 ``Finite dimensional systems with random external fields and neutrino propagation in fluctuating media,''
 Phys.\ Rev.\ D {\bf 59}, 073001 (1999)
 [hep-ph/9807361].
 
  
  
\bibitem{Loreti:1995ae} 
  F.~N.~Loreti, Y.~Z.~Qian, G.~M.~Fuller and A.~B.~Balantekin,
  ``Effects of random density fluctuations on matter enhanced neutrino flavor transitions in supernovae and implications for supernova dynamics and nucleosynthesis,''
  Phys.\ Rev.\ D {\bf 52}, 6664 (1995)
  [astro-ph/9508106].
  
\bibitem{Fogli:2006xy} 
  G.~L.~Fogli, E.~Lisi, A.~Mirizzi and D.~Montanino,
  ``Damping of supernova neutrino transitions in stochastic shock-wave density profiles,''
  JCAP {\bf 0606}, 012 (2006)
  [hep-ph/0603033].
  
 \bibitem{Choubey:2007ga}
 S.~Choubey, N.~P.~Harries and G.~G.~Ross,
 ``Turbulent supernova shock waves and the sterile neutrino signature in megaton water detectors,''
 Phys.\ Rev.\ D {\bf 76}, 073013 (2007)
 [hep-ph/0703092 [HEP-PH]].

\bibitem{Benatti:2004hn} 
  F.~Benatti and R.~Floreanini,
  ``Dissipative neutrino oscillations in randomly fluctuating matter,''
  Phys.\ Rev.\ D {\bf 71}, 013003 (2005)
  [hep-ph/0412311].
  
\bibitem{Friedland:2006ta} 
  A.~Friedland and A.~Gruzinov,
  ``Neutrino signatures of supernova turbulence,''
  astro-ph/0607244.
  
\bibitem{Kneller:2010sc} 
  J.~P.~Kneller and C.~Volpe,
  ``Turbulence effects on supernova neutrinos,''
  Phys.\ Rev.\ D {\bf 82}, 123004 (2010)
  [arXiv:1006.0913 [hep-ph]].
  
\bibitem{random}
A.~J.~Majda and P.~R.~Kramer,  ``Simplified models for turbulent diffusion: Theory, numerical modelling, and physical phenomena,''
Phys.\ Rep. \ {\bf 314} 237 (1999).

\bibitem{Kneller:2013ska} 
  J.~P.~Kneller and A.~W.~Mauney,
  ``The consequences of large $\theta_{13}$ for the turbulence signatures in supernova neutrinos,''
  arXiv:1302.3825 [hep-ph].
  
\bibitem{Lund:2013uta} 
  T.~Lund and J.~P.~Kneller,
  ``Combining collective, MSW, and turbulence effects in supernova neutrino flavor evolution,''
  Phys.\ Rev.\ D {\bf 88}, 023008 (2013)
  [arXiv:1304.6372 [astro-ph.HE]].
 
 
  
   \bibitem{rashba}
   T.~Rashba, ``Effects of random density fluctuations on Supernovae and solar neutrinos.'' Talk at
   the \emph{Workshop on Next generation Nucleon decay and Neutrino detectors, NNN 2006}, Seattle 21--23
September 2006. Slides available at 
{\tt http://neutrino.phys.washington.edu/nnn06/slides/
rashba\_supernovae\_density\_fluctuations\_v2.pdf}.  


\bibitem{Dolence:2013} 
  J.C.~Dolence, A.~Burrows, J.W.~Murphy and J.~Nordhaus,
  ``Dimensional dependence of the hydrodynamics of core-collapse supernovae,''
  Astrophys.\ J.\ {\bf 765}, 110 (2013)
  [arXiv:1210.5241].


\bibitem{Couch:2013b}
  S.M.~Couch and E.P.~O'Connor,
  ``High-resolution three-dimensional simulations of core-collapse supernovae in multiple progenitors,''
  [arXiv:1310.5728].


\bibitem{Woosley:1988at} 
  S.~E.~Woosley,
  ``SN1987A: After the peak,''
  Astrophys.\ J.\  {\bf 330}, 218 (1988).

\bibitem{Arcones:2007}
A.~Arcones, H.-T.~Janka and L.~Scheck,
  ``Nucleosynthesis-relevant conditions in neutrino-driven supernova outflows. Spherically symmetric hydrodynamic simulations,''
  Astron.\ Astrophys.\  {\bf 467}, 1227 (2007)
  [astro-ph/0612582].


\bibitem{Kuo:1989qe} 
  T.~-K.~Kuo and J.~T.~Pantaleone,
  ``Neutrino Oscillations in Matter,''
  Rev.\ Mod.\ Phys.\  {\bf 61}, 937 (1989).

\bibitem{Fogli:2012ua} 
  G.~L.~Fogli, E.~Lisi, A.~Marrone, D.~Montanino, A.~Palazzo and A.~M.~Rotunno,
  ``Global analysis of neutrino masses, mixings and phases: entering the era of leptonic CP violation searches,''
  Phys.\ Rev.\ D {\bf 86}, 013012 (2012)
  [arXiv:1205.5254 [hep-ph]].
  
  \bibitem{kraichnan}
  R.~Kraichnan, ``Inertial ranges in two-dimensional turbulence,'' Phys.\ Fluids {\bf 10}, 1417 (1967);
R.~Kraichnan, ``Inertial-range transfer in two- and three-dimensional turbulence,'' Journ.\ Fluid Mech.\ {\bf 47}, 525 (1971);
R.~Kraichnan and D.~Montgomery, ``Two-dimensional turbulence,'' Rep.\ Prog.\ Phys.\ {\bf 43}, 547 (1980).
  
 \bibitem{boffetta}
 G.~Boffetta and  R.~E.~Ecke, ``Two-Dimensional Turbulence,''
Annual Review of Fluid Mechanics {\bf 44}, 427 (2012). 
  
  
\bibitem{Dasgupta:2005wn} 
  B.~Dasgupta and A.~Dighe,
  ``Phase effects in neutrino conversions during a supernova shock wave,''
  Phys.\ Rev.\ D {\bf 75}, 093002 (2007)
  [hep-ph/0510219].
  





\bibitem{Arnett:2011}
  W.D.~Arnett and C.~Meakin,
  ``Toward realistic progenitors of core-collapse supernovae,''
  Astrophys.\ J.\ {\bf 733}, 78 (2011)
  [arXiv:1101.5646].


\bibitem{Couch:2013a}
  S.M.~Couch and C.D.~Ott,
  ``Revival of the stalled core-collapse supernova shock triggered by precollapse asphericity 
in the progenitor star,''
  [arXiv:1309.2632].
  
  
\bibitem{Kneller:2013lka} 
  J.~P.~Kneller and A.~W.~Mauney,
  ``Does the finite size of the proto-neutron star preclude supernova neutrino flavor scintillation due to turbulence?,''
  Phys.\ Rev.\ D {\bf 88}, no. 4, 045020 (2013).
  
\bibitem{Fischer:2009af} 
  T.~Fischer, S.~C.~Whitehouse, A.~Mezzacappa, F.~-K.~Thielemann and M.~Liebendorfer,
  ``Protoneutron star evolution and the neutrino driven wind in general relativistic neutrino radiation hydrodynamics simulations,''
  Astron.\ Astrophys.\  {\bf 517}, A80 (2010)
  [arXiv:0908.1871 [astro-ph.HE]].
  
  
\end{thebibliography}
\end{document}